\documentstyle[preprint,fleqn,aps]{revtex}

\begin{document}

\hoffset -1cm

\draft
\preprint{TPR-95-24}

\title{ Relativistic Kinetic Equations for Electromagnetic,\\
        Scalar and Pseudoscalar Interactions}

\author{Pengfei Zhuang\cite{home}}

\address{Gesellschaft f\"ur Schwerionenforschung, Theory Group, \\
         P.O.Box 110552, D-64220 Darmstadt, Germany}

\author{Ulrich Heinz}

\address{Institut f\"ur Theoretische Physik, Universit\"at Regensburg,\\
         D-93040 Regensburg, Germany }

\date{\today}

\maketitle

\begin{abstract}

We derive the kinetic equations for both the covariant and equal-time Wigner
functions of Dirac particles with electromagnetic, scalar and pseudoscalar
interactions. We emphasize the constraint equations for the spinor components
in the equal-time formulation.

\end{abstract}



One of the recent progresses in transport theory \cite{EH,He} is the
establishment of transport equations for spinor \cite{BGR} and scalar
\cite{BGG} equal-time Wigner functions with abelian gauge interaction.
The main advantage of the equal-time formulation lies in the fact that
the initial value of the equal-time Wigner function can be directly
obtained from the corresponding initial field operators, since there
is only one time scale in the equal-time formulation. Therefore some
quantum problems such as pair production \cite{Sch,KESCM} in strong
electric fields have so far only be solved in the equal-time
formulation \cite{BGR,BE}. The advantages of the covariant
formulation, on the other hand, are explicit Lorentz invariance and
the feature that the kinetic equations naturally split into a
transport equation of Vlasov-Boltzmann type and a generalized
mass-shell condition \cite{EH,He} which makes explicit the off-shell
effects generated by the collisions in the system. In Ref. \cite{ZH}
we discussed the relationship between the covariant and equal-time
kinetic equations. By taking the energy average of the covariant
equations we obtained both transport and constraint equations for
the equal-time Wigner functions in scalar and spinor electrodynamics.

Recently, Shin and Rafelski \cite{SR} discussed a system which, in
addition to the electromagnetic vector interaction, included also
scalar and pseudoscalar (strong) interactions. For the system defined
by the Lagrangian
 \begin{eqnarray}
 \label{Lagrangian}
    {\cal L} &=& \bar\psi (i\gamma^\mu\partial_\mu -e\gamma^\mu A_\mu
    -m +g_\sigma\sigma +ig_\pi\pi\gamma_5 )\psi-{1\over 4}F_{\mu\nu}
             F^{\mu\nu}
 \nonumber \\
     & & +{1\over 2}(\partial^\mu \sigma\partial_\mu \sigma+\partial^\mu
             \pi \partial_\mu \pi)-{\cal U}_M(\sigma,\pi)\, ,
 \end{eqnarray}
they derived a generalized group of transport equations for the spinor
components of the equal-time Wigner function in the mean field
approximation. Here $A_\mu, \sigma$ and $\pi$ are electromagnetic,
scalar and pseudoscalar fields, respectively, $g_\sigma$ and $g_\pi$
are the scalar and pseudoscalar coupling constants, and ${\cal
U}_M(\sigma,\pi)$ is the mesonic self-interaction potential.
In this short paper, we first study the full covariant
kinetic equations for such a system and then derive from them
transport and constraint equations also for the equal-time Wigner
function, using our recently introduced energy averaging method. We
thereby supplement the equal-time transport equations derived in
Ref.~\cite{SR} by a set of non-trivial constraint equations. The
importance of the latter is emphasized both for the semiclassical
limit and for the general quantum dynamics.

In the following we restrict ourselves to the kinetic equations for
the spinor Wigner function in the mean field approximation, by
replacing the field operators $A_\mu, \sigma$ and $\pi$ by their mean
values. Such a Hartree approximation can be justified for sufficiently
strong fields and has so far been used in most applications of quantum
transport theory. Since the meson potential ${\cal U}_M$ does not
explicitly appear in the kinetic equations for the spinor Wigner
function (it only enters the selfconsistency equations for the scalar
and pseudoscalar mean fields), the discussions below holds universally
for all potentials ${\cal U}_M(\sigma,\pi)$.

The covariant spinor Wigner function $W_4(x,p)$ is the ensemble
average of the Wigner operator $\hat W_4(x,p)$ which is the
4-dimensional Wigner transformation of the covariant density operator
$\hat \Phi_4(x,y)$:
 \begin{eqnarray}
 \label{W4}
   W_4(x,p) &=& \int d^4 y\, e^{ipy} \langle \hat \Phi_4(x,y) \rangle
 \nonumber\\
            &=& \int d^4y\, e^{ipy} \left \langle \psi(x+{y\over 2} )
                \exp\left[ ie\int^{1\over 2}_{-{1\over 2}}ds
                     A(x+sy){\cdot}y \right]
                     \bar \psi(x-{y\over 2}) \right\rangle \, .
 \end{eqnarray}

Calculating the first-order derivatives of the covariant density
matrix $\hat \Phi_4(x,y)$ with respect to $x$ and $y$ and using the
Dirac equation
 \begin{equation}
 \label{Dirac}
   i\gamma_\mu\Big(\partial^\mu_x+ieA^\mu(x)\Big)\psi(x)=
   m\psi(x)-g_\sigma\sigma(x)\psi(x)-ig_\pi\pi(x)\gamma^5\psi(x)\,
 \end{equation}
one obtains an evolution equation for $\hat \Phi_4(x,y)$. After
Fourier transforming with respect to the 4 components of the relative
coordinate $y$ one obtains for arbitrary external fields $A_\mu,
\sigma$ and $\pi$ the generalized VGE \cite{VGE} equation
 \begin{equation}
 \label{VGE}
   [\gamma^\mu K_\mu +\gamma^5 K_5 -M]\hat W_4 = 0\, ,
 \end{equation}
where now
 \begin{eqnarray}
 \label{4operator}
   && K_\mu = \Pi_\mu +{i\hbar\over 2}D_\mu\, ,
\nonumber \\
   && K_5 = \Pi_5+iD_5\, ,
 \nonumber \\
   && M = M_1 + iM_2\, ,
 \nonumber \\
   && \Pi_\mu(x,p)
      = p_\mu - ie\hbar \int ^{1\over 2}_{-{1\over 2}} ds\, s\,
        F_{\mu\nu}(x-i\hbar s\partial_p) \, \partial_p^\nu \, ,
 \nonumber\\
   && D_\mu(x,p)
      = \partial_\mu - e\int ^{1\over 2}_{-{1\over 2}} ds\,
        F_{\mu\nu} (x-i\hbar s\partial_p)\, \partial_p^\nu \, ,
 \nonumber \\
   && \Pi_5(x,p) = g_\pi\sin\left({\hbar\over 2}\Delta\right) \pi(x)\, ,
 \nonumber\\
   && D_5(x,p) = g_\pi \cos \left({\hbar\over 2} \Delta\right) \pi(x)\, ,
 \nonumber \\
   && M_1(x,p) = m-g_\sigma\cos \left({\hbar\over 2}\Delta\right) \sigma(x)\, ,
 \nonumber \\
   && M_2(x,p) = g_\sigma\sin \left({\hbar\over 2}\Delta\right) \sigma(x)\, ,
 \nonumber\\
   && \Delta = \partial_x{\cdot}\partial_p\, .
 \end{eqnarray}
For the discussion of the semiclassical expansion below we have
explicitly given all factors of $\hbar$. If one neglects in these
equations the abelian gauge field tensor and all terms of second or
higher order in $\hbar$ one recovers the results recently obtained by
Florkowski et al.~\cite{FHKN}.

Inserting the spinor decomposition \cite{VGE}
 \begin{equation}
 \label{W4spin}
    \!\!\!\!\!\!\!\!\!\!\!\!
    W_4(x,p) = {1\over 4} \left[ F(x,p) + i \gamma_5 P(x,p)
             + \gamma_\mu V^\mu(x,p) + \gamma_\mu \gamma_5 A^\mu(x,p)
             + {1\over 2}\sigma_{\mu\nu} S^{\mu\nu}(x,p)\right] \, ,
 \end{equation}
we split the complex equation (\ref{VGE}) into 10 real equations for
the spinor components:
 \begin{eqnarray}
 \label{VGEspin}
    &&\Pi^\mu V_\mu -D_5 P= M_1 F \, ,
 \nonumber \\
    && \hbar D^\mu A_\mu +2D_5 F= 2M_1 P \, ,
 \nonumber \\
    &&\Pi_\mu F +{\hbar\over 2}D^\nu S_{\mu\nu}-\Pi_5A_\mu=
      M_1 V_\mu \, ,
 \nonumber \\
    &&-\hbar D_\mu P+\epsilon_{\mu\nu\sigma\rho}\Pi^\nu S^{\sigma\rho}
      -2\Pi_5V_\mu = 2M_1 A_\mu \, ,
 \nonumber \\
    &&{\hbar\over 2} (D_\mu V_\nu-D_\nu V_\mu)
      + \epsilon_{\mu\nu\sigma\rho}\Pi^\sigma A^\rho
      -{1\over 2}\epsilon_{\mu\nu\sigma\rho}D_5S^{\sigma\rho}
      = M_1 S_{\mu\nu} \, ,
 \nonumber\\
    && \hbar D^\mu V_\mu +2\Pi_5 P = 2M_2 F \, ,
 \nonumber \\
    &&\Pi^\mu A_\mu +\Pi_5 F = -M_2 P \, ,
 \nonumber \\
    &&{\hbar\over 2} D_\mu F - \Pi^\nu S_{\nu\mu} -D_5A_\mu =
      M_2 V_\mu\, ,
 \nonumber \\
    &&\Pi_\mu  P - D_5 V_\mu +{\hbar\over 4}\epsilon_{\mu\nu\sigma\rho}
      D^\nu S^{\sigma\rho} = M_2 A_\mu\, ,
 \nonumber \\
    &&\Pi_\mu V_\nu-\Pi_\nu V_\mu - {\hbar\over 2}\epsilon_{\mu\nu\sigma\rho}
      D^\sigma A^\rho - {1\over 2}\epsilon_{\mu\nu\sigma\rho}\Pi_5
      S^{\sigma\rho} = M_2 S_{\mu\nu}\, .
 \end{eqnarray}
Again the corresponding results up to first order in $\hbar$ can be
found in Ref. \cite{FHKN} if we do not consider the abelian gauge interaction.

Following the treatment of Ref. \cite{ZH}, we write the equal-time
Wigner function $W_3(x,{\bf p})$ as the energy average of the
covariant Wigner function $W_4(x,p)$ and decompose it in a similar
way \cite{BGR} into its spinor components:
 \begin{eqnarray}
 \label{W3}
   \!\!\!\!\!\!\!\!\!\!
   W_3(x,{\bf p}) \!\!&=&\!\! \int d^3{\bf y} e^{-i{\bf p}{\cdot}{\bf y}}
                \left\langle \psi\left({\bf x}+{{\bf y}\over 2}, t \right)
                \exp\left[ ie\int^{1\over 2}_{-{1\over 2}}ds
                    {\bf A}({\bf x}+s{\bf y},t){\cdot}{\bf y} \right]
                \psi^\dagger\left({\bf x}-{{\bf y} \over 2}, t\right)
                \right\rangle
 \nonumber\\
       &=&\!\! \int dp_0\, W_4(x,p)\, \gamma_0
 \nonumber\\
       &=&\!\! {1\over 4}\Bigl[ f_0(x,{\bf p})
                        + \gamma_5 f_1(x,{\bf p})
                        -i \gamma_0 \gamma_5 f_2(x,{\bf p})
                        + \gamma_0 f_3(x,{\bf p})\nonumber\\
       && + \gamma_5 \gamma_0{\bf \gamma}{\cdot}{\bf g}_0(x,{\bf p})
          + \gamma_0 {\bf \gamma}{\cdot}{\bf g}_1(x,{\bf p})
          - i{\bf \gamma}{\cdot}{\bf g}_2(x,{\bf p})
          - \gamma_5 {\bf \gamma}{\cdot}{\bf g}_3(x,{\bf p})
          \Bigr]\, .
 \end{eqnarray}
The second line in (\ref{W3}) suggests that the equal-time kinetic
equations may be obtained by taking the energy average of the
corresponding covariant equations. As pointed out in Ref. \cite{ZH}
and seen below, this procedure yields additional information on the
equal-time Wigner function, in the form of additional constraint
equations which are lost in the procedure used in Ref. \cite{SR} which
is based on a 3-dimensional Wigner transformation of the equations of
motion for the equal-time density matrix.

The energy average of the equations (\ref{VGEspin}) gives rise to kinetic
equations for the spinor components $f_i$ and ${\bf g}_i\ (i=0,1,2,3)$ of
the equal-time Wigner function. In direct analogy to Ref. \cite{ZH}
one obtains a set of transport equations which are identical to the
ones derived in \cite{SR},
 \begin{eqnarray}
 \label{transport}
    \hbar(D_t f_0 + {\bf D}\cdot {\bf g}_1)
    &=&-2g_\sigma \sigma_o f_3\!+\! 2g_\pi\pi_o f_2\, ,
 \nonumber \\
    \hbar(D_t f_1 + {\bf D}\cdot {\bf g}_0)
    &=& 2g_\sigma\sigma_e f_2\!+\!2g_\pi\pi_e f_3 -2mf_2\, ,
 \nonumber\\
    \hbar(D_t f_2 + 2{\bf \Pi}\cdot {\bf g}_3)
    &=& - 2g_\sigma\sigma_e f_1\!+\! 2g_\pi\pi_o f_0 +2mf_1 \, ,
 \nonumber\\
    \hbar(D_t f_3 - 2{\bf \Pi}\cdot{\bf g}_2)
    &=& -2g_\sigma\sigma_o f_0\!-\! 2g_\pi\pi_e f_1\, ,
 \nonumber\\
    \hbar(D_t {\bf g}_0+{\bf D} f_1)- 2{\bf \Pi}\!\times\!{\bf g}_1
    &=& -2g_\sigma\sigma_o{\bf g}_3 \!+\!2g_\pi\pi_o{\bf g}_2\, ,
 \nonumber\\
    \hbar(D_t {\bf g}_1 + {\bf D} f_0)-2{\bf \Pi}\!\times\! {\bf g}_0
    &=& 2g_\sigma\sigma_e{\bf g}_2 \!+\!2g_\pi\pi_e{\bf g}_3
       -2m{\bf g}_2\, ,
 \nonumber\\
    \hbar(D_t {\bf g}_2 + {\bf D}\!\times\!{\bf g}_3) + 2{\bf \Pi} f_3
    &=& -2g_\sigma\sigma_e{\bf g}_1 \!+\!2g_\pi\pi_o{\bf g}_0
        +2m{\bf g}_1 \, ,
 \nonumber\\
    \hbar(D_t {\bf g}_3-{\bf D}\!\times\!{\bf g}_2) -2{\bf \Pi} f_2
    &=& -2g_\sigma\sigma_o{\bf g}_0 \!-\!2g_\pi\pi_e{\bf g}_1\, ,
 \end{eqnarray}
as well as a set of additional constraint equations:
 \begin{eqnarray}
 \label{constraint1}
    &&\int dp_0 p_0 F - {1\over 2}\hbar {\bf D}\cdot {\bf g}_2
      + \tilde\Pi_0 f_3 -g_\pi \pi_o f_1 = (m-g_\sigma \sigma_e) f_0 \, ,
 \nonumber \\
    &&\int dp_0 p_0 P + {1\over 2}\hbar {\bf D}\cdot {\bf g}_3
      + \tilde\Pi_0 f_2 -g_\pi \pi_e f_0 - g_\sigma \sigma_o f_1 = 0 \, ,
 \nonumber \\
    &&\int dp_0 p_0 V_0 -{\bf \Pi}\cdot {\bf g}_1 + \tilde\Pi_0 f_0
      -g_\pi \pi_e f_2 = (m-g_\sigma \sigma_e) f_3 \, ,
 \nonumber \\
    &&\int dp_0 p_0 {\bf V} - {1\over 2}\hbar {\bf D}\times {\bf g}_0
      - {\bf \Pi} f_0 + \tilde\Pi_0 {\bf g}_1 +g_\pi \pi_o {\bf g}_3
      +g_\sigma \sigma_o {\bf g}_2 = 0 \, ,
 \nonumber\\
    &&\int dp_0 p_0 A_0 +{\bf \Pi}\cdot {\bf g}_0 - \tilde\Pi_0 f_1
      - g_\pi \pi_o f_3 - g_\sigma \sigma_o f_2 = 0 \, ,
 \nonumber \\
    &&\int dp_0 p_0 {\bf A} + {1\over 2}\hbar {\bf D}\times {\bf g}_1
      + {\bf \Pi} f_1 - \tilde\Pi_0 {\bf g}_0 + g_\pi \pi_e {\bf g}_2
      = -(m-g_\sigma \sigma_e) {\bf g}_3 \, ,
 \nonumber \\
    &&\int dp_0 p_0 S^{0i}{\bf e}_i - {1\over 2}\hbar {\bf D} f_3
      + {\bf \Pi}\times {\bf g}_3 - \tilde\Pi_0 {\bf g}_2
      + g_\pi \pi_e {\bf g}_0 + g_\sigma \sigma_o {\bf g}_1 = 0 \, ,
 \nonumber \\
    &&\int dp_0 p_0 S_{jk} \epsilon^{jki}{\bf e}_i
      - \hbar{\bf D} f_2 +2{\bf \Pi}\times {\bf g}_2
      + 2 \tilde\Pi_0 {\bf g}_3 -2 g_\pi \pi_o {\bf g}_1
      =2 (m-g_\sigma \sigma_e) {\bf g}_0 \, .
 \end{eqnarray}
Here we defined the following 3-dimensional operators
 \begin{eqnarray}
 \label{3operator1}
    D_t(x,{\bf p})
    &=& \partial_t + e \int^{1\over 2}_{-{1\over 2}} ds\,
        {\bf E}({\bf x}+is\hbar{\bf \nabla}_p, t){\cdot}{\bf \nabla}_p
    \, ,
 \nonumber \\
    {\bf D}(x,{\bf p})
    &=& {\bf \nabla} + e \int^{1\over 2}_{-{1\over 2}}ds\,
    {\bf B}({\bf x}+is\hbar{\bf \nabla}_p, t) \times {\bf \nabla}_p
    \, ,
 \nonumber \\
    {\bf \Pi}(x,{\bf p})
    &=& {\bf p}-i e \hbar \int^{1\over 2}_{-{1\over 2}}ds\, s\,
    {\bf B}({\bf x} + is\hbar{\bf \nabla}_p, t) \times {\bf \nabla}_p
    \, ,
 \nonumber\\
    \tilde \Pi_0 (x,{\bf p})
    &=& i e \hbar
    \int^{1\over 2}_{-{1\over 2}}ds\, s\, {\bf E}({\bf x}
         + i s \hbar {\bf \nabla}_p,t){\cdot}{\bf \nabla}_p
    \, ,
 \nonumber\\
    \sigma_e(x,{\bf p})
    &=& \cos({\hbar\over 2}{\bf \nabla}\cdot {\bf \nabla}_p) \sigma(x)
    \, ,
 \nonumber \\
    \sigma_o(x,{\bf p})
    &=& \sin({\hbar\over 2}{\bf \nabla}\cdot {\bf \nabla}_p) \sigma(x)
    \, ,
 \nonumber\\
    \pi_e(x,{\bf p})
    &=&  \cos({\hbar\over 2}{\bf \nabla}\cdot {\bf \nabla}_p) \pi(x)
    \, ,
 \nonumber\\
    \pi_o(x,{\bf p})
    &=&  \sin({\hbar\over 2}{\bf \nabla}\cdot {\bf \nabla}_p) \pi(x)\, ,
 \end{eqnarray}
which correspond to the covariant operators (\ref{4operator}). ${\bf
E}$ and ${\bf B}$ are the electric and magnetic fields, respectively.

We now discuss the relevance of the additional constraint equations
(\ref{constraint1}), both in the classical limit and for genuine
quantum situations. In the classical limit, the covariant kinetic
equations have solutions \cite{ZH} of the form $W_4^\pm(x,p) =
\tilde W_4^\pm(x,{\bf p})\,\delta(p_0\mp E_p)$. Here $E_p$ is still an
arbitrary function of $p$; its concrete form will be seen momentarily
as a result of the constraint equations. To zeroth order in $\hbar$,
the kinetic equations (\ref{transport}) and (\ref{constraint1}) lead
to the following constraints for the classical spinor components and
the quasiparticle energy $E_p$:
 \begin{eqnarray}
 \label{constraintC1}
  && f_1^\pm = \pm {{\bf p}{\cdot}{\bf g}_0^\pm\over E_p}\, ,
     \nonumber\\
  && f_2^\pm = \pm {V_\pi f_0^\pm \over E_p}\, ,
     \nonumber\\
  && f_3^\pm = \pm {(m-V_\sigma) f_0^\pm\over E_p}\, ,
     \nonumber\\
  && {\bf g}_1^\pm = \pm {{\bf p}f_0^\pm\over E_p}\, ,
     \nonumber\\
  && {\bf g}_2^\pm = {{\bf p}\times {\bf g}_0^\pm +
     V_\pi {\bf g}_3^\pm \over m-V_\sigma}\, ,
     \nonumber\\
  && {\bf g}_3^\pm = \pm {E_p^2 (m-V_\sigma) {\bf g}_0^\pm
                          -(m-V_\sigma){\bf p}({\bf p}{\cdot}{\bf g}_0^\pm)
                          \mp E_p V_\pi{\bf p}\times {\bf g}_0^\pm
     \over E_p (m^*)^2}\, ,
     \nonumber\\
  && E_p^2 = (m^*)^2+{\bf p}^2\, .
\end{eqnarray}
Here we defined
 \begin{equation}
 \label{3operator2}
   V_\sigma = g_\sigma \sigma\, , \qquad
   V_\pi = g_\pi \pi \, , \qquad
   (m^*)^2 = (m-V_\sigma)^2+V_\pi^2 \, .
 \end{equation}
The classical transport equations for the charge and spin densities
are seen to originate from the equation (\ref{transport}) at first
order in $\hbar$:
 \begin{eqnarray}
 \label{transportC1}
    d_t f_0^\pm + {\bf d}{\cdot}{\bf g}_1^\pm
    &=& ({\bf F}_\sigma{\cdot} {\bf \nabla}_p)f_3^\pm
      - ({\bf F}_\pi {\cdot} {\bf \nabla}_p)f_2^\pm\, ,
 \nonumber\\
    d_t {\bf g}_3^\pm - {\bf d}\times {\bf g}_2^\pm
    &=& ({\bf F}_\sigma{\cdot}{\bf \nabla}_p){\bf g}_0^\pm
        - {V_\pi\over m-V_\sigma}({\bf F}_\pi
        {\cdot} {\bf \nabla}_p){\bf g}_0^\pm
 \nonumber\\
    && -{{\bf p}\over m-V_\sigma}(d_t f_1^\pm +{\bf d}{\cdot}{\bf g}_0^\pm)
     - {V_\pi\over m-V_\sigma}(d_t {\bf g}_2^\pm +{\bf d}
     \times {\bf g}_3^\pm)\, ,
  \end{eqnarray}
where
 \begin{eqnarray}
 \label{3operator3}
  && d_t = \partial_t + e{\bf E}{\cdot}{\bf \nabla}_p\, ,
     \qquad  {\bf d} = {\bf \nabla} + e{\bf B}\times {\bf \nabla}_p\, ,
 \nonumber\\
  && {\bf F}_\sigma = -g_\sigma {\bf \nabla}\sigma\, ,
     \qquad  {\bf F}_\pi = -g_\pi {\bf \nabla}\pi\, .
 \end{eqnarray}
Introducing the particle and antiparticle charge densities $f$ and
$\bar f$ and spin densities ${\bf g}$ and $\bar {\bf g}$ through
 \begin{eqnarray}
 \label{3operator4}
   && f(x,{\bf p}) = f_0^+(x,{\bf p})\, ,\qquad
      \bar f(x,{\bf p}) = f_0^-(x,-{\bf p})\, ,
 \nonumber\\
   && {\bf g}(x,{\bf p}) = {\bf g}_0^+(x,{\bf p})\, ,\qquad
       \bar {\bf g}(x,{\bf p}) = {\bf g}_0^-(x,-{\bf p})\, ,
 \end{eqnarray}
and using the constraints (\ref{constraintC1}), one derives from
(\ref{transportC1}) a Vlasov-type transport equation for $f$,
 \begin{equation}
 \label{transportC2}
     \partial_t f+ ({\bf v}{\cdot}{\bf \nabla})f
     + \left(e{\bf E}+e{\bf v}\times {\bf B}
     - {(m-V_\sigma){\bf F}_\sigma - V_\pi {\bf F}_\pi\over E_p}
       \right)
     {\cdot}{\bf \nabla}_p f = 0\, ,
 \end{equation}
and a similar equation for the spin density ${\bf g}$:
 \begin{eqnarray}
   &&\!\!\!\!
     \partial_t {\bf g}+ ({\bf v}{\cdot}{\bf \nabla}){\bf g}
     + \left(e{\bf E}+e{\bf v}\times {\bf B}
     - {(m-V_\sigma){\bf F}_\sigma - V_\pi {\bf F}_\pi\over E_p}
       \right)
     {\cdot}{\bf \nabla}_p {\bf g}
 \nonumber\\
   &&\!\!\!\!
     = {e\over E_p^2} \Bigg(({\bf p}{\cdot}{\bf g}){\bf E}
                           -({\bf E}{\cdot}{\bf p})
       {\bf g}\Bigg) - {e\over E_p}{\bf B}\times {\bf g}
 \nonumber\\
   &&+{(m-V_\sigma){\bf F}_\pi +V_\pi {\bf F}_\sigma \over (m^*)^2}\times
     \Bigg({\bf g} -{({\bf p}{\cdot}{\bf g})\over E_p^2}{\bf p}\Bigg)
     +{(m-V_\sigma){\bf F}_\sigma-V_\pi{\bf F}_\pi\over E_p (m^*)^2}
     \times ({\bf p}\times {\bf g})
 \nonumber\\
   &&+{(m-V_\sigma) \partial_t V_\pi +V_\pi \partial_t V_\sigma
     \over (m-V_\sigma)^2 (m^*)^2}V_\pi\Bigg(E_p {\bf g}-{({\bf p}
     {\cdot}{\bf g})\over E_p}{\bf p}
     -{V_\pi\over m-V_\sigma}{\bf p}\times {\bf g}\Bigg)
 \nonumber\\
   &&-{m^* \partial_t m^*\over E_p^4} \Bigg(({\bf p}{\cdot}{\bf g}){\bf p}
     +(m^*)^2{\bf g}\Bigg)+{\partial_t V_\sigma \over (m-V_\sigma)E_p}
     \Bigg({V_\pi\over m-V_\sigma}{\bf p}
     \times {\bf g}-{(m^*)^2\over E_p}{\bf g}\Bigg)\, .
 \label{transportC2a}
 \end{eqnarray}
Here ${\bf v} = {{\bf p}\over E_p}$ is the velocity of the classical
transport flow. The corresponding equations for the antiparticle
densities $\bar f$ and $\bar {\bf g}$ are obtained by changing the
sign of the electric charge, $e \to -e$. The spin evolution equation
(\ref{transportC2a}) is the 3-dimensional phase-space version of a
generalized Bargmann-Michel-Telegdi (BMT) equation \cite{BMT,VGE}
which describes spin precession in external electromagnetic, scalar
and pseudoscalar fields. The first two terms of the r.h.s. which
couple the spin density to the electromagnetic field have already been
given in \cite{ZH}. The remaining terms describe spin-precession
effects caused by the coupling to the scalar and pseudoscalar fields.
They are the 3-dimensional version of the much more compact covariant
expressions derived in \cite{FHKN}.

It is well known \cite{VGE,FHKN} that the covariant spin equation can
be greatly simplified by introducing the covariant ``spin up" and
``spin down" distribution functions
 \begin{equation}
 \label{spin4}
  F_{\pm s}(x,p) = {F(x,p)\over m-V_\sigma}\pm {S_\mu(x,p)A^\mu (x,p)\over
  m^*(x)}\, ,
 \end{equation}
where $S_\mu=A_\mu/(-A{\cdot}A)^{1/2}$ is the covariant and normalized
spin phase-space density. The kinetic equations for the analogous
equal-time distribution functions $f_{\pm s}(x,{\bf p}) = E_p \int
dp_0\, F_{\pm s}(x,p)\,\delta (p_0 - E_p)$ and $\bar f_{\pm s}(x,{\bf
p}) = E_p \int dp_0\, F_{\pm s}(x,-p)\,\delta (p_0 + E_p)$ decouple
into scalar equations of the form (\ref{transportC2}) also obeyed by
the charge density:
 \begin{equation}
 \label{spin3}
   \partial_t f_{\pm s}+ ({\bf v}{\cdot}{\bf \nabla})f_{\pm s}
   + \left(e{\bf E}+e{\bf v}\times {\bf B}
   - {(m-V_\sigma){\bf F}_\sigma - V_\pi {\bf F}_\pi\over E_p}
   \right){\cdot}{\bf \nabla}_p f_{\pm s} = 0\, ,
 \end{equation}
and a similar one for the antiparticle distribution $\bar f_{\pm s}$
obtained through $e \to -e$.

Although the energy averaging method and the procedure of
Ref.~\cite{SR} lead to the same Vlasov transport equations for $f$ and
$\bar f$, we would like to point out an essential difference: Since
the equal-time constraint equations (\ref{constraint1}) do not
disappear even in the classical limit, they reduce (together with the
transport equations (\ref{transport}) to order $\hbar^0$) the number
of independent spinor components from 8 in \cite{SR} (namely $f_0,
f_3, {\bf g}_0$ and ${\bf g}_3$) to 4 here (namely the charge density
$f_0$ and the spin density ${\bf g}_0$). Furthermore, the form of the
quasiparticle dispersion relation $E_p$ need not be guessed as in
\cite{SR}, but follows from the constraints. In the treatment of
Ref.~\cite{SR}, the equations for the charge density $f_0$ and the
mass density $f_3$ remained coupled, and the Vlasov-type transport
equation (\ref{transportC2}) was only obtained after a suitable
redefinition of the classical distribution functions by hand (see
equations (57) and (58) in \cite{SR}). Our constraint equations
(\ref{constraint1}) turn these redefinition equations into identities.

For genuine quantum problems like pair production, one must consider
the non-perturpative (in $\hbar$) form of the constraint equations
(\ref{constraint1}). To get a closed set of equations for the spinor
components of the equal-time Wigner function, one eliminates the first
$p_0$-moments $\int dp_0\, p_0\, W_4(x,p)$ from (\ref{constraint1}) by
combining them with the first $p_0$-moments of those covariant
equations in (\ref{VGEspin}) which gave rise to the equal-time
transport equations (\ref{transport}). The procedure is entirely
analogous to the one detailed in Ref. \cite{ZH}, and one obtains the
constraints
 \begin{eqnarray}
 \label {constraint2}
    &&L f_0+{\bf M}\cdot {\bf g}_1
      -{\bf F}_{\sigma_o}\cdot {\bf g}_2
      -{\bf F}_{\pi_o}\cdot {\bf g}_3=0 \, ,
 \nonumber \\
    &&L f_1+{\bf M}\cdot {\bf g}_0
      -({\bf F}_{\sigma_e}+{\bf F}_\pi)\cdot {\bf g}_2
      +({\bf F}_{\pi_e}+{\bf F}_\sigma)\cdot {\bf g}_3=0 \, ,
 \nonumber \\
    &&L f_2+2{\bf N}\cdot {\bf g}_3
      -({\bf F}_{\pi_e}-{\bf F}_\pi)\cdot {\bf g}_1
      -{\bf F}_{\sigma_o}\cdot {\bf g}_0=0 \, ,
 \nonumber \\
    &&L f_3-2{\bf N}\cdot {\bf g}_2
      +({\bf F}_{\sigma_e}-{\bf F}_\sigma)\cdot {\bf g}_1
      -{\bf F}_{\pi_o}\cdot {\bf g}_0=0 \, ,
 \\
    &&L {\bf g}_0 -{\bf M} f_1 -2{\bf N}\times {\bf g}_1
      -{\bf F}_{\sigma_o} f_2
      -{\bf F}_{\pi_o} f_3
      -({\bf F}_{\pi_e}-{\bf F}_\pi)\times {\bf g}_3
      -({\bf F}_{\sigma_e}-{\bf F}_\sigma)\times{\bf g}_2= 0 \, ,
 \nonumber \\
    &&L {\bf g}_1 -{\bf M} f_0 -2{\bf N}\times {\bf g}_0
      -({\bf F}_{\pi_e}-{\bf F}_\pi) f_2
      +({\bf F}_{\sigma_e}-{\bf F}_\sigma) f_3
      +{\bf F}_{\sigma_o}\times {\bf g}_3
      -{\bf F}_{\pi_o}\times {\bf g}_2=0 \, ,
 \nonumber \\
    &&L {\bf g}_2 +{\bf M}\times {\bf g}_3 -2{\bf N} f_3
      +{\bf F}_{\sigma_o} f_0
      +{\bf F}_{\pi_e} f_1
      -({\bf F}_{\sigma_e}-{\bf F}_\sigma)\times {\bf g}_0
      +{\bf F}_{\pi_o}\times {\bf g}_1= 0 \, ,
 \nonumber \\
    &&L {\bf g}_3 -{\bf M}\times {\bf g}_2 +2{\bf N} f_2
      +{\bf F}_{\pi_o} f_0
      -({\bf F}_{\sigma_e}-{\bf F}_\sigma) f_1
      -({\bf F}_{\pi_e}-{\bf F}_\pi)\times {\bf g}_0
      -{\bf F}_{\sigma_o}\times {\bf g}_1= 0 \, ,
 \nonumber
 \end{eqnarray}
with
 \begin{eqnarray}
 \label{3operator5}
    && \!\!\!\!\!\!\!\!\!\!\!
      L(x,{\bf p}) = i e \int^{1\over 2}_{-{1\over 2}} ds\, s\,
      \Big({\bf \nabla}\times {\bf B}({\bf x}+is{\bf \nabla}_p,t)
      \Big) \cdot {\bf \nabla}_p \, ,
 \nonumber \\
    && \!\!\!\!\!\!\!\!\!\!\!
      {\bf M}(x,{\bf p}) = i e \int^{1\over 2}_{-{1\over 2}} ds\, s\,
      {\bf \nabla} \Big({\bf E}({\bf x}+is{\bf \nabla}_p,t){\cdot}{\bf
          \nabla}_p\Big) + e \int^{1\over 2}_{-{1\over 2}} ds\,
      \Big({\bf E}({\bf x}+is{\bf \nabla}_p,t) - {\bf E}(x)\Big) \, ,
 \nonumber \\
    && \!\!\!\!\!\!\!\!\!\!\!
       {\bf N}(x,{\bf p}) = {1\over 4} e \int^{1\over 2}_{-{1\over 2}}ds\,
       ({\bf \nabla}_p{\cdot}{\bf \nabla})\,
       {\bf E}({\bf x}+is{\bf \nabla}_p,t) \nonumber\\
    &&\ \ \ \ \ \ \ \ \ +e \int^{1\over 2}_{-{1\over 2}}ds\,
       s^2\Big(\partial_t {\bf B}({\bf x}+is{\bf
       \nabla}_p,t)\Big)\times {\bf \nabla}_p +ie \int^{1\over 2}_{-
       {1\over 2}}ds s{\bf E}({\bf x}+is{\bf \nabla}_p,t)\, ,
 \nonumber\\
    && {\bf F}_{\sigma_o} = -g_\sigma {\bf \nabla} \sigma_o\, ,
 \nonumber\\
    && {\bf F}_{\sigma_e} = -g_\sigma {\bf \nabla} \sigma_e\, ,
 \nonumber\\
    && {\bf F}_{\pi_o} = -g_\pi {\bf \nabla} \pi_o\, ,
 \nonumber\\
    && {\bf F}_{\pi_e} = -g_\pi {\bf \nabla} \pi_e\, .
 \end{eqnarray}
These constraints, which are closely associated with the quantum
corrections to the classical mass-shell condition, hold for arbitrary
external fields. Only for homogeneous fields, the spatial derivatives
of the electric, magnetic, scalar and pseudoscalar fields vanish, and
therefore the constraint equations (\ref{constraint2}) disappear
identically.

These results extend our work in \cite{ZH} where we studied only
electromagnetic interactions. They also complement the recent studies
of Refs.~\cite{FHKN} and \cite{SR} by generalizing the former to all
orders in $\hbar$ and supplementing the latter with the necessary
additional constraints. The generalization to scalar and pseudoscalar
interactions is relevant for the development of a chiral kinetic
theory with possible applications to the non-equilibrium dynamics in
relativistic heavy ion collisions. Eq.~(\ref{Lagrangian}) has the
generic form of an effective chiral Lagrangian for low-energy hadronic
physics; suitable choices for the meson potential ${\cal U}_M$
correspond to different popular effective chiral models. For example,
replacing the second line in Eq. (\ref{Lagrangian}) by
$C(\sigma^2+\pi^2)$ \cite{FHKN} yields the semi-bosonized version of
the Nambu--Jona-Lasinio model \cite{NJL}. On the other hand, the meson
potential\ ${\cal U}_M\sim (\sigma^2+\pi^2-v^2)^2$, with $v$ as a
parameter, represents the original linear $\sigma$ model \cite{sigma}.
These two models are widely used to describe low energy hadronic
physics and the chiral phase transition at high temperatures and
densities. Therefore, the study of the kinetic equations in both
covariant and equal-time formulations for the Lagrangian
(\ref{Lagrangian}) is helpful for an understanding of the dynamical
consequences of the chiral properties of QCD.

We should not close without repeating from Ref. \cite{ZH} an important
cautionary remark. The energy averaging procedure elucidates that the
kinetic equations for the equal-time Wigner function $W_3(x,{\bf p}) =
\int dp_0\, W_4(x,p)$ form only the lowest level of a coupled hierarchy
of equations for the energy moments of the covariant Wigner function,
$\int dp_0 \, p_0^n\, W_4(x,p),\ n=0,1,2,\dots$ The full hierarchy can be
obtained by taking the energy average of the appropriate $p_0$-moments
of the generalized VGE equation (\ref{VGE}). Its properties are
presently under study in the context of the simpler case of scalar QED
\cite{ZHa}. The connection between certain truncation schemes for this
hierarchy of equal-time moment equations and the well-known gradient
expansion in the covariant approach \cite{EH,He,FHKN} is an
interesting subject for further research.

\acknowledgments

P.Z. wishes to thank the GSI for a fellowship. This work was supported
in part by BMBF and DFG.



\begin{thebibliography}{99}

  \bibitem[*]{home}  On leave from Hua-Zhong Normal University, Wuhan,
               China.
  \bibitem{EH}   {H.-Th. Elze and U. Heinz, Phys. Rep. {\bf 183}, 81
                  (1989).}
  \bibitem{He}   {P. Henning, Phys. Rep. {\bf 253}, 263 (1995).}
  \bibitem{BGR}  {I. Bialynicki-Birula, P. Gornicki and J. Rafelski,
                  Phys. Rev. D{\bf 44}, 1825 (1991).}
  \bibitem{BGG}  {C. Best, P. Gornicki and W. Greiner, Ann. Phys.
                  (N.Y.) {\bf 225}, 169 (1993).}
  \bibitem{Sch}  {J. Schwinger, Phys. Rev. {\bf 82}, 664 (1951).}
  \bibitem{KESCM}{Y. Kluger, J. M. Eisenberg, B. Svetitsky, F. Cooper,
                  and E. Mottola, Phys. Rev. Lett. {\bf 67}, 2427
                  (1991); and Phys. Rev. D{\bf 45}, 4659 (1992).}
  \bibitem{BE}   {C. Best and J. M. Eisenberg, Phys. Rev. D{\bf 47},
                  4639 (1993).}
  \bibitem{ZH}   {P. Zhuang and U. Heinz, Ann. Phys. (N.Y.), in press.}
  \bibitem{SR}   {G. R. Shin and J. Rafelski, Ann. Phys. (N.Y.), in press.}
  \bibitem{VGE}  {D. Vasak, M. Gyulassy, and H.-Th. Elze, Ann. Phys.
                  (N.Y.) {\bf 173}, 462 (1987).}
  \bibitem{FHKN} {W. Florkowski, J. H\"ufner, S.P. Klevansky, and L.
                  Neise, {\it Chirally invariant transport equations
                  for quark matter}, Heidelberg preprint HD-TVP-95-5
                  (hep-ph/9505407), Ann. Phys., in press.}
  \bibitem{BMT}  {V. Bargmann, L. Michel, and V. Telegdi, Phys. Rev. Lett.
                  {\bf 2}, 435 (1959);\\
                  see also U.Heinz, Ann. Phys. (N.Y.) {\bf 161}, 48 (1985).}
  \bibitem{NJL}  For example, see {S.P.Klevansky, Rev. Mod. Phys.
                 {\bf 64}, 649 (1992).}
  \bibitem{sigma}{M. Gell-Mann and M. L\'{e}vy, Nuovo Cim. {\bf 16},
                  705 (1960).}
  \bibitem{ZHa}   {P. Zhuang and U. Heinz, {\it Equal-time kinetic
                   equations for scalar electrodynamics}, Regensburg
                   preprint TPR-95-25, in preparation.}

\end{thebibliography}
\end{document}